\documentclass[lettersize,journal]{IEEEtran}
\usepackage{amsmath,amsfonts}
\usepackage{algorithm}
\usepackage{enumitem}
\usepackage{booktabs}
\usepackage{booktabs, makecell}
\usepackage{array}
\usepackage[caption=false,font=normalsize,labelfont=sf,textfont=sf]{subfig}
\usepackage{textcomp}
\usepackage{stfloats}
\usepackage{url}
\usepackage{verbatim}
\usepackage{graphicx}
\usepackage{cite}
\usepackage{amsmath,amssymb,amsfonts}  
\usepackage{booktabs}  
\usepackage{ragged2e}  
\usepackage{float}  
\usepackage{cite}  
\usepackage{multirow}  
\usepackage{hyperref}  
\usepackage{caption}  
\usepackage{subcaption}  
\usepackage{svg}  
\usepackage{algorithm}  
\usepackage{algpseudocode}  
\usepackage{enumitem}  
\usepackage{braket}  
\usepackage{array}  
\usepackage{caption}
\usepackage{comment}
\usepackage{graphicx}  
\begin{document}



\title{Quantum Skyshield: Quantum Key Distribution and Post-Quantum Authentication for Low-Altitude Wireless Networks in Adverse Skies}

\author{Zeeshan Kaleem,~\IEEEmembership{Senior Member,~IEEE}, Misha Urooj Khan, Ahmad Suleman, Waqas Khalid, \\Kai-Kit Wong,~\IEEEmembership{Fellow,~IEEE}, Chau Yuen,~\IEEEmembership{Fellow,~IEEE}    %
\thanks{This work is supported and funded by Cybersecurity Research and Innovation Pioneers Grants Initiative with grant number CRPG-25-3065. (*Corresponding author: Zeeshan Kaleem, e-mail: zeeshankaleem@gmail.com).

Zeeshan Kaleem is with the Department of Computer Engineering and the Interdisciplinary Research Center for Smart Mobility and Logistics, King Fahd University of Petroleum \& Minerals (KFUPM), Dhahran 31261, Saudi Arabia (e-mail: zeeshankaleem@gmail.com)

Misha Urooj Khan is with the European Organization for Nuclear Research, CERN, Switzerland, and chairperson of CRD (e-mail: misha.urooj.khan@cern.ch)

Ahmad Suleman is affiliated with AITeC, National Center for Physics (NCP), Pakistan, and serves as Vice-chairperson of the Community of Research and Development (CRD) (e-mail: engineersuleman118@gmail.com)

Waqas Khalid is with the Institute of Industrial Technology, Korea University, Sejong 30019, South Korea (e-mail: waqas283@gmail.com)

K. K. Wong is affiliated with the Department of Electronic and Electrical Engineering, University College London, Torrington Place, WC1E 7JE, United Kingdom, and also with Yonsei Frontier Lab, Yonsei University, Seoul, Korea. (e-mail: kai-kit.wong@ucl.ac.uk)

Chau Yuen is with the School of Electrical and Electronics Engineering,
Nanyang Technological University, Singapore (email: chau.yuen@ntu.edu.sg).}}

\maketitle
\begin{abstract}
Recently, low-altitude wireless networks (LAWNs) have emerged as a critical backbone for supporting the low-altitude economy, particularly with the densification of unmanned aerial vehicles (UAVs) and high-altitude platforms (HAPs). To meet growing data demands, some LAWN deployments incorporate free-space optical (FSO) links, which offer exceptional bandwidth and beam directivity. However, without strong security measures in place, both conventional radio frequency channels and FSO beams remain vulnerable to interception and spoofing—and FSO in particular can suffer from turbulence, misalignment, and weather-related attenuation.
To address these challenges in the quantum era, quantum-secure architecture called \textit{Quantum Skyshield} is proposed to enable reliable communication between the base transceiver station (BTS) and LAWN. The proposed design integrates BB84 quantum key distribution (QKD) with post-quantum authentication mechanisms. Simulation results confirm the reliable generation of a 128-bit symmetric key when the quantum bit error rate (QBER) remains below the threshold of 11\%. Authentication is enforced using Lamport one-time signatures and hash-based message authentication codes (HMAC) to ensure message integrity. A Grover-inspired threat detection mechanism identifies anomalies with up to 89\% probability in a single iteration, enabling real-time trust evaluation. Lastly, future research challenges have also been identified and discussed to guide further development in this area.
\end{abstract}


\section{Introduction}
\IEEEPARstart{Q}{uantum} information processing is set to transform communication technologies by enabling ultra-secure data exchange, anonymous networking, distributed computing, and precision sensing. These capabilities are crucial across various sectors, including defense, healthcare, finance, and the metaverse \cite{sharma2021}. 

Low-altitude wireless networks (LAWNs), comprising unmanned aerial vehicles (UAVs) and high-altitude platforms (HAPs) operating in the lower atmosphere, are emerging as a vital component of non-terrestrial networks (NTNs). These networks support critical applications such as disaster response, environmental sensing, surveillance, and low-latency broadband access in under-served regions \cite{alshaer2021}. As LAWN deployments expand, the limitations of traditional radio frequency (RF) communication—namely jamming, eavesdropping, and spectrum congestion—are becoming increasingly pronounced. In response, free-space optical (FSO) communication has gained attention for its high bandwidth, narrow beamwidth, and inherent resistance to interception. However, its effectiveness in LAWN scenarios is constrained by challenges such as atmospheric turbulence, beam misalignment due to platform mobility, and weather-induced attenuation \cite{sinha2023,guo2022}.

To secure LAWN links, QKD over FSO has emerged as a viable strategy. Protocols like Bennett–Brassard 1984 (BB84) have demonstrated potential in enabling provably secure key exchange over optical paths between LAWN nodes and base transceiver stations (BTS) \cite{nguyen2023}. However, several limitations persist. First, physical-layer disturbances, including beam wander in LAWNs or ionospheric interference in LEO satellites, increase quantum bit error rate (QBER), making it difficult to distinguish between channel noise and active eavesdropping \cite{sinha2023}. Second, traditional key reconciliation and error correction procedures are resource-intensive and often infeasible for dynamic and energy-constrained LAWN nodes. 
\begin{figure*}[h]
    \centering
    \includegraphics[width=1.05\textwidth]{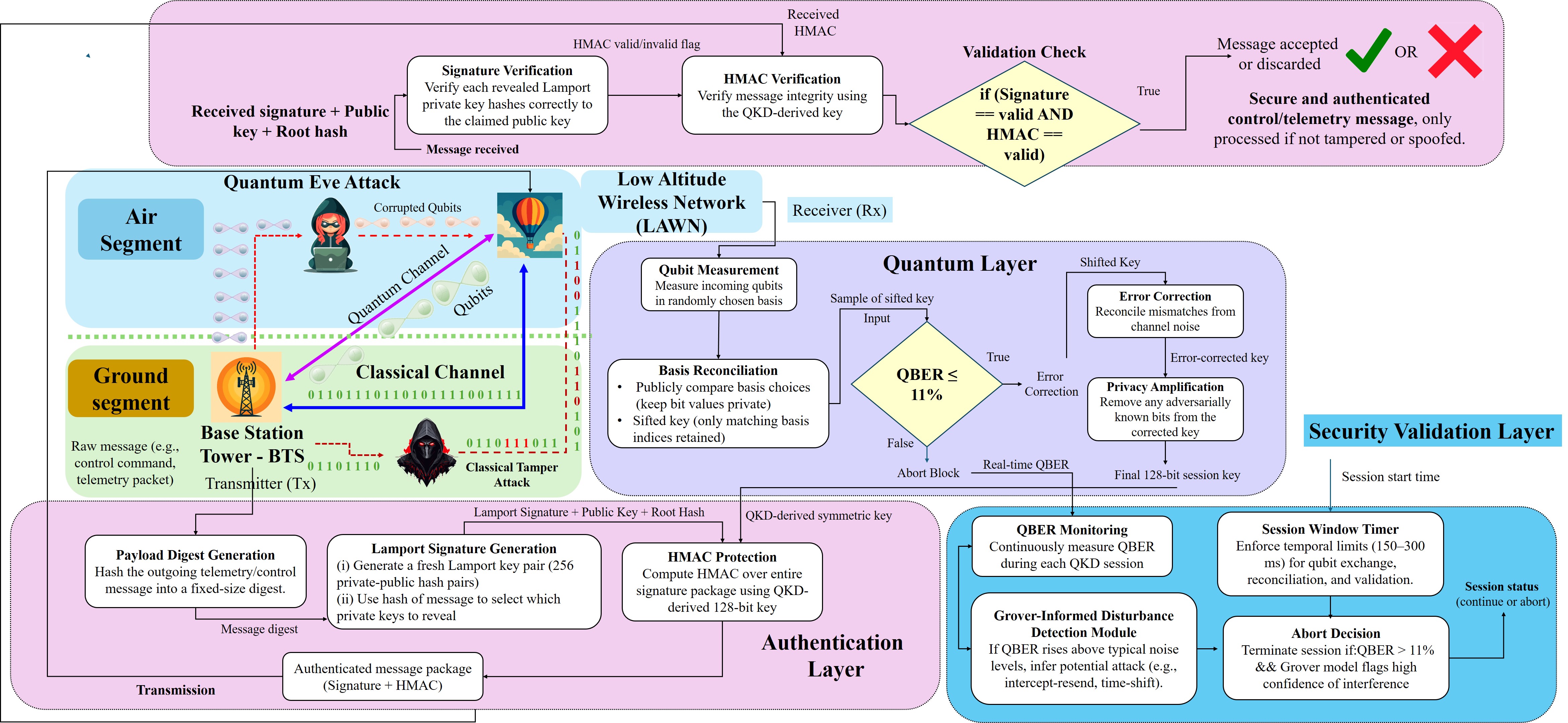}
    \caption{Proposed Quantum Skyshield: Post-Quantum Secure QKD Architecture for BTS-LAWN Communication.}
    \label{Fig:1}
\vspace{-0.5cm}
\end{figure*}
Many existing systems still rely on quantum-insecure schemes such as Rivest Shamir Adleman (RSA) and Elliptic Curve Digital Signature Algorithm (ECDSA), which are susceptible to Shor’s algorithm \cite{sharma2024}. While post-quantum cryptographic (PQC) solutions, such as CRYSTALS-Dilithium and Kyber, provide stronger guarantees, their high computational overhead limits their adoption on lightweight LAWN platforms \cite{lu2022,ramabadran2023}. Though hash-based signature schemes offer lightweight alternatives \cite{guo2022}, integration with QKD-secured LAWN remains limited. Furthermore, most current frameworks focus narrowly on quantum key exchange, often neglecting holistic telemetry protection and adversarial detection mechanisms essential for robust communication \cite{alallaq2024}. These gaps highlight the urgent need for a scalable, quantum-resilient architecture designed explicitly for LAWN nodes operating under resource and environmental constraints. 

To address these challenges, we propose a layered cryptographic architecture, \textit{Quantum skyshield}, that supports secure QKD and post-quantum authentication across LAWN. Our contributions include:
\begin{enumerate}
  \item Existing QKD designs in \cite{sinha2023,gupta2023,chehimi2024} ignored real-world impairments like atmospheric turbulence, misalignment, and orbital jitter. We incorporate a Gamma-Gamma turbulence-based FSO model and implement BB84 QKD with complete post-processing, including basis reconciliation, QBER estimation, error correction, and privacy amplification. Moreover, sessions are aborted if QBER exceeds a defined security threshold, ensuring robust key integrity.

  \item While prior works \cite{alshaer2022,chehimi2024} treat QBER as a static metric, we embed an active Grover’s algorithm-based anomaly detection mechanism. This transforms fluctuating QBER levels into adversarial indicators, enabling real-time threat detection and session trust evaluation across LAWN links.

  \item We replaced RSA and ECDSA with Lamport one-time signatures (OTS), a hash-based authentication scheme requiring only lightweight operations. Paired with hash-based message authentication codes (HMACs) derived from QKD keys, our method enables secure communication without relying on pre-established trust channels or resource-intensive lattice schemes \cite{sharma2024,guo2022}.

  \item Recognizing that even with QKD, classical telemetry can be compromised, we integrate quantum-derived symmetric keys and hash-based authentication to secure command and control traffic. This hybrid model ensures end-to-end message integrity even in GPS-denied, RF-contested, or space-jammed environments \cite{alallaq2024}.
\end{enumerate}

\section{Quantum Skyshield: Post-Quantum Secure QKD Architecture for BTS-LAWN Communication}
Unlike classical encryption protocols that often expose metadata and rely on pre-established trust, our architecture uses QKD-derived keys to authenticate control signals and anonymize node identities during transmission across LAWN links. By coupling symmetric key-based authentication with hash-based digital signatures, the system ensures both message integrity and sender untraceability, addressing a critical gap in NTN. This integrated approach lays the foundation for scalable, privacy-preserving quantum communication in non-terrestrial environments, as shown in Fig.~\ref{Fig:1}. Currently, we have implemented the architecture for LAWN-to-BTS links, which also serves as an air-to-ground NTN use case. With minimal changes, such as minor loss adjustments and synchronization tweaks, it can be extended to broader NTN scenarios, including LEO-to-ground, and LAWN-to-LAWN links. These use cases include secure satellite control, disaster recovery swarms, and quantum mesh LAWN networks.

\subsubsection{FSO Channel Modeling Layer}
This layer captures the stochastic nature of FSO links under real-world operational conditions. The model incorporates both deterministic and probabilistic effects across variable propagation distances as illustrated in Fig.~\ref{Fig:2}(a). Atmospheric attenuation, weather-specific visibility loss (e.g., fog, rain, snow), and turbulence-induced beam distortions are characterized using a Gamma–Gamma distribution. Additionally, misalignment and pointing jitter are modeled as dynamic angular deviations that can lead to partial or total packet loss. The cumulative effect is quantified as a total channel gain, determining whether individual quantum or classical packets are successfully received or dropped.

\subsubsection{Authentication Layer}
The system utilizes a Lamport OTS scheme, a one-time signature method that involves applying cryptographic hash functions to random secret values. Each Lamport key pair consists of 256 pairs of private-public values, and to sign a message, the signer reveals one private value from each pair depending on the hash of the message. This eliminates the need for computationally intensive operations, such as modular exponentiation or lattice constructions. Its advantage lies in its simplicity, low computational demand, and resistance to quantum attacks, making it ideal for lightweight platforms like LAWNs. It is selected here specifically because it enables post-quantum-secure message authentication using only hash functions, which are far less demanding than lattice-based alternatives and well-suited to the limited resources of airborne nodes, rooted in hash-based cryptography, to ensure payload authenticity and post-quantum robustness. The BTS generates a fresh Lamport key pair for each transmission, signs the payload digest, and sends the resulting signature with its full public key and a root hash. To safeguard against classical tampering, an HMAC is computed over the entire signature package using the 128-bit QKD-derived key. 

At the LAWN, signature verification involves checking that each revealed private-key value hashes correctly to the associated public key, and confirming the HMAC tag to validate message integrity. This layered approach ensures that only authentic, untampered messages from a verified source are accepted, even in the presence of a quantum adversary or a compromised classical channel, as shown in Fig.~\ref{Fig:2}(b).

\subsubsection{Quantum Layer}
The BB84 protocol, implemented over the modeled FSO link, forms the core of the proposed architecture. In each transmission session, the BTS (Alice) generates a stream of single-photon qubits. Each qubit is prepared in one of two non-orthogonal bases: rectilinear $\{0^\circ, 90^\circ\}$ or diagonal $\{45^\circ, 135^\circ\}$, with bit values assigned randomly. These photons are transmitted through the FSO channel to the LAWN (Bob), which measures each qubit using a randomly chosen basis. Due to atmospheric impairments and the mobility of LAWNs, not all photons arrive intact; some are lost or misaligned. Those successfully received are measured and recorded. 

After the quantum transmission, Alice and Bob engage in a basis reconciliation process over the classical channel. They publicly disclose their respective basis choices for each photon while keeping the bit values private. Bits corresponding to matching bases are retained, forming the sifted key. From this sifted key, a random sample is publicly revealed to calculate the QBER. If the QBER is within the acceptable threshold ($\leq 11\%$), the remaining key undergoes classical post-processing. This includes error correction to reconcile discrepancies due to channel noise, ensuring bit-wise agreement. 

The output is a 128-bit symmetric session key that guarantees information-theoretic secrecy. This key is then passed to the classical layer, where it is used for authenticating LAWN telemetry, control commands, and HMAC generation, as depicted in Fig.~\ref{Fig:2}(c). This layered mechanism ensures that even under degraded atmospheric conditions or partial photon loss, the integrity of secure key generation and transmission is preserved.
\subsubsection{Security Validation Layer}
The architecture integrates a security validation mechanism inspired by Grover’s quantum search algorithm, which enables adversaries to search symmetric key spaces quadratically faster than classical brute-force methods. This speed advantage makes brute-force attacks against symmetric encryption and authentication primitives more feasible in post-quantum environments, necessitating tighter thresholds for anomaly detection. While Grover’s algorithm is not executed directly, its implications inform the system’s adversarial threat model and reinforce the importance of monitoring even subtle increases in the QBER. In this context, any significant deviation from expected QBER levels may indicate active quantum-layer attacks, such as photon number splitting, intercept-resend, or time-shift attacks. 

To implement this, the system includes a Grover-informed disturbance detection module that continually monitors QBER trends. For high-mobility aerial links, each QKD session is constrained to a typical operational window of 150–300 milliseconds, within which photon exchange, basis reconciliation, and QBER estimation must be completed. If QBER exceeds the secure threshold (11\%), the session is flagged for potential compromise, and key extraction is aborted. In parallel, classical-layer security is enforced through message integrity verification. The LAWN validates the authenticity of each incoming message by checking the Lamport one-time signature and the HMAC derived from the quantum key. If either verification fails, the message is rejected, and a possible spoofing, replay, or tampering incident is logged. 

\begin{figure*}[h]
    \centering
    (a) \includegraphics[width=.3\textwidth]{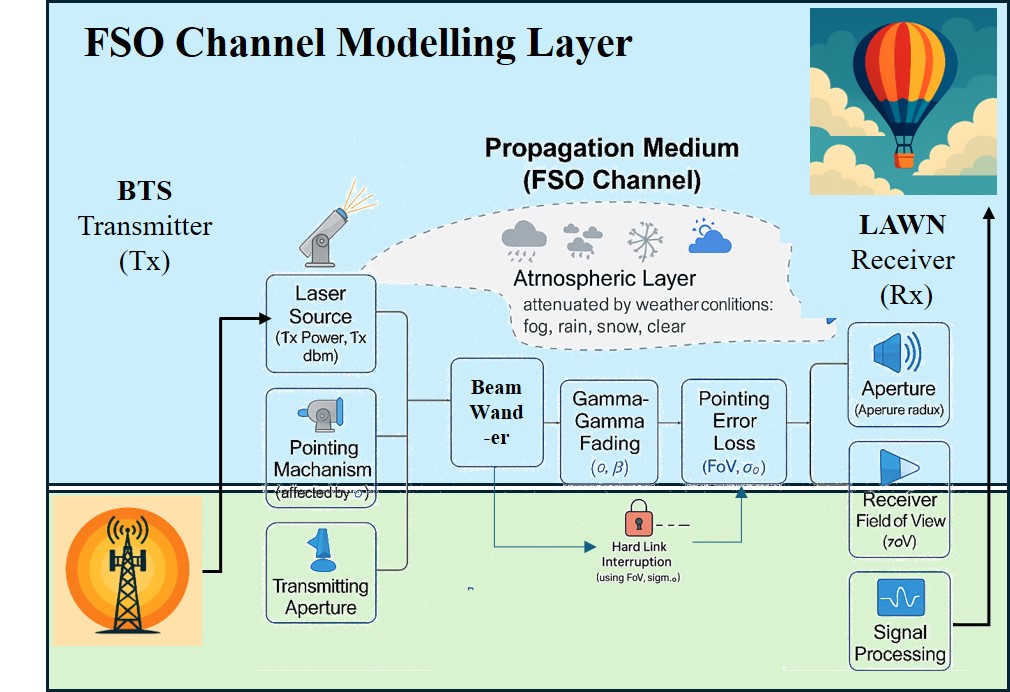} 
    (b) \includegraphics[width=.55\textwidth]{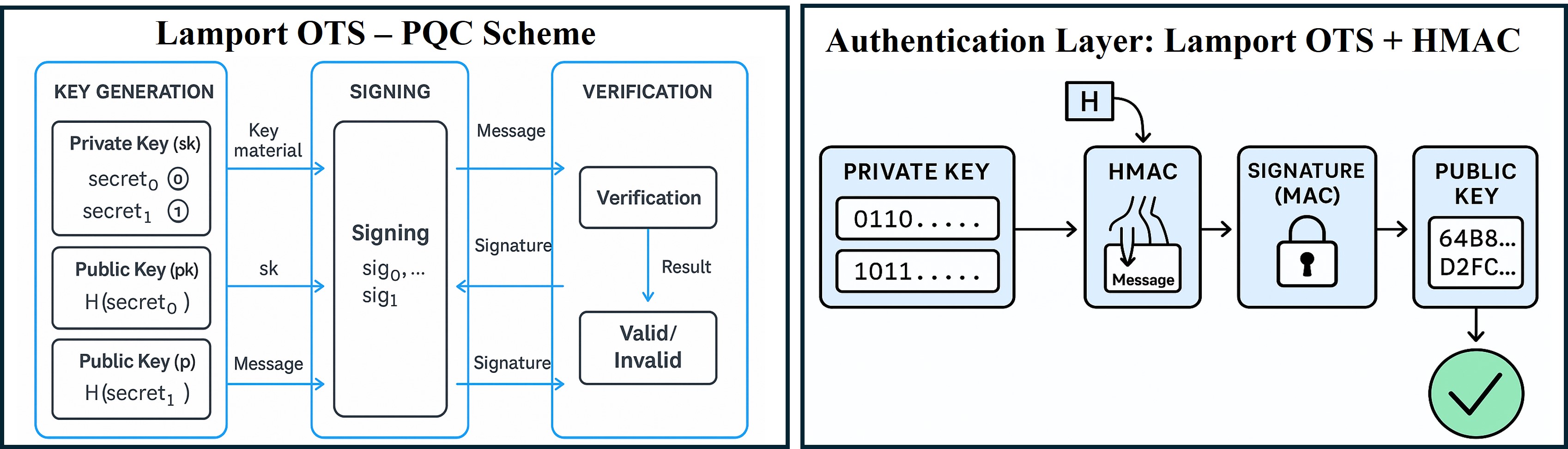} \\
    (c) \includegraphics[width=.97\textwidth]{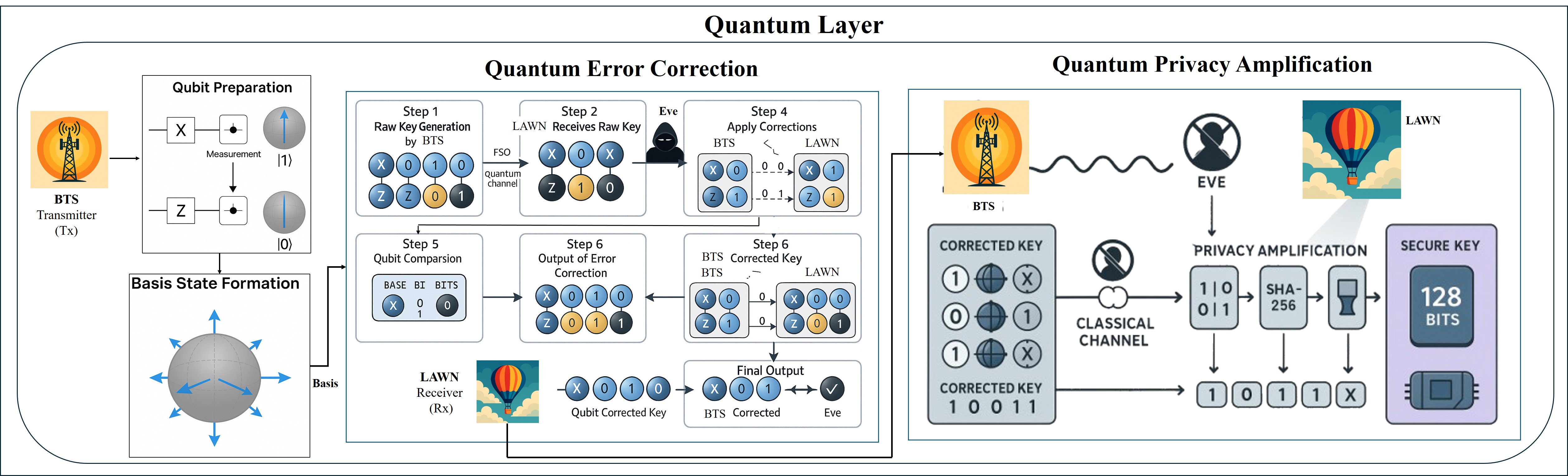}\\
    (d) \includegraphics[width=.90\textwidth]{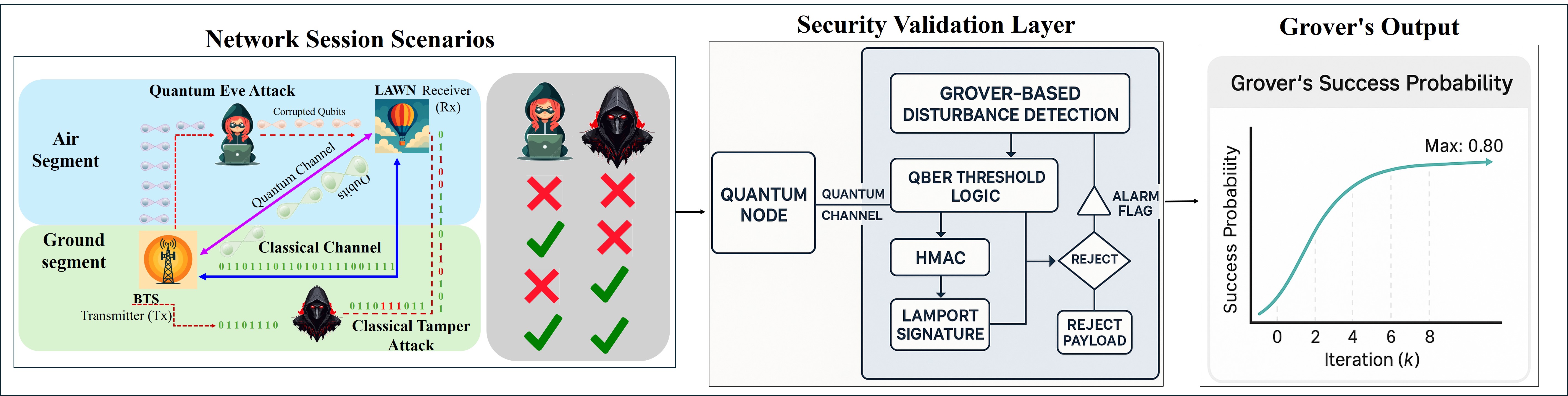}
    \caption{(a) FSO Channel Modelling Layer (b) Authentication Layer: Lamport OTS + HMAC (c) Quantum Layer (d) Security Validation Layer.}
    \label{Fig:2}
\vspace{-0.5cm}
\end{figure*}

As illustrated in Fig.~\ref{Fig:2}(d), this dual-layer validation ensures that both quantum and classical threats are detected early, preserving the trustworthiness of the overall communication session.

\subsubsection{System Workflow and Evaluation}
The complete framework is executed sequentially in a system-level simulation environment that captures realistic LAWN-BTS FSO dynamics. Link parameters, including atmospheric turbulence, beam divergence, and pointing error, directly affect both qubit delivery and classical packet reliability. Each simulation begins with BB84-based QKD, where photon transmission, basis reconciliation, and QBER estimation are performed over the modeled FSO link. If the observed QBER remains within the secure threshold ($\leq 11\%$), the system extracts a valid 128-bit symmetric key through error correction and privacy amplification. Once the key is established, the classical layer is invoked. A synthetic telemetry payload is signed using the Lamport OTS, and the entire authentication package is protected via HMAC derived from the QKD key. This emulates secure telemetry exchange between the BTS and LAWN.

The evaluation includes two adversarial scenarios to assess system robustness: (i) an active quantum-layer attacker performing intercept-resend operations, which raises QBER and may force session abortion; and (ii) a classical-layer attacker attempting payload tampering, which is detected by either an invalid Lamport signature or a failed HMAC verification. Both scenarios are crucial for demonstrating the system's multi-layered defense capabilities. Output metrics include the sifted key length, QBER levels, final session key acceptance or rejection, signature verification status, and HMAC validation results. These metrics are essential for quantifying the operational integrity of each subsystem and the overall architecture's effectiveness. Together, they offer comprehensive insights into the feasibility, resilience, and deployability of quantum-resilient LAWN communication in contested or degraded environments.

\begin{figure*}[h]
    \centering
    (a)\includegraphics[width=.46\textwidth]{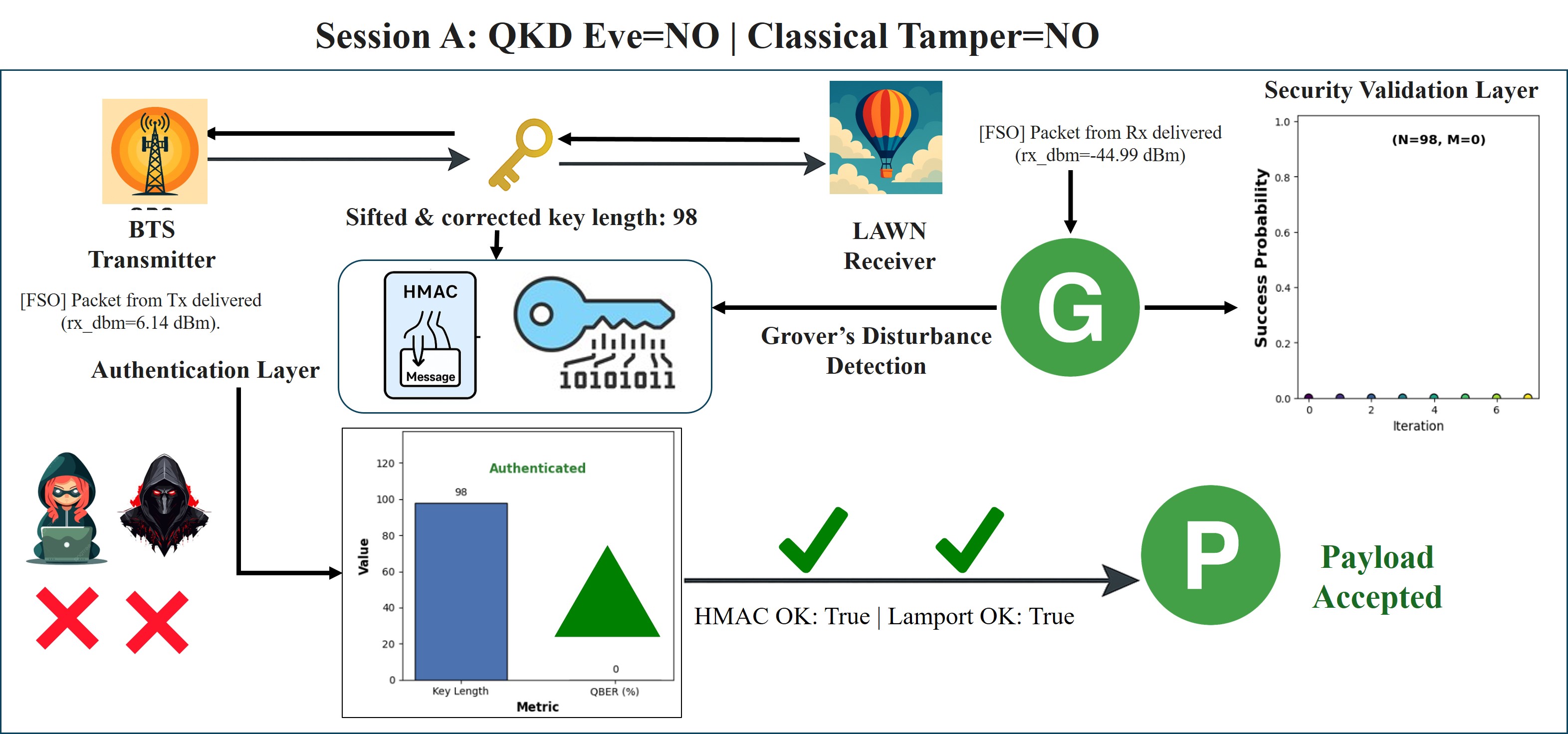} 
    (b)\includegraphics[width=.46\textwidth]{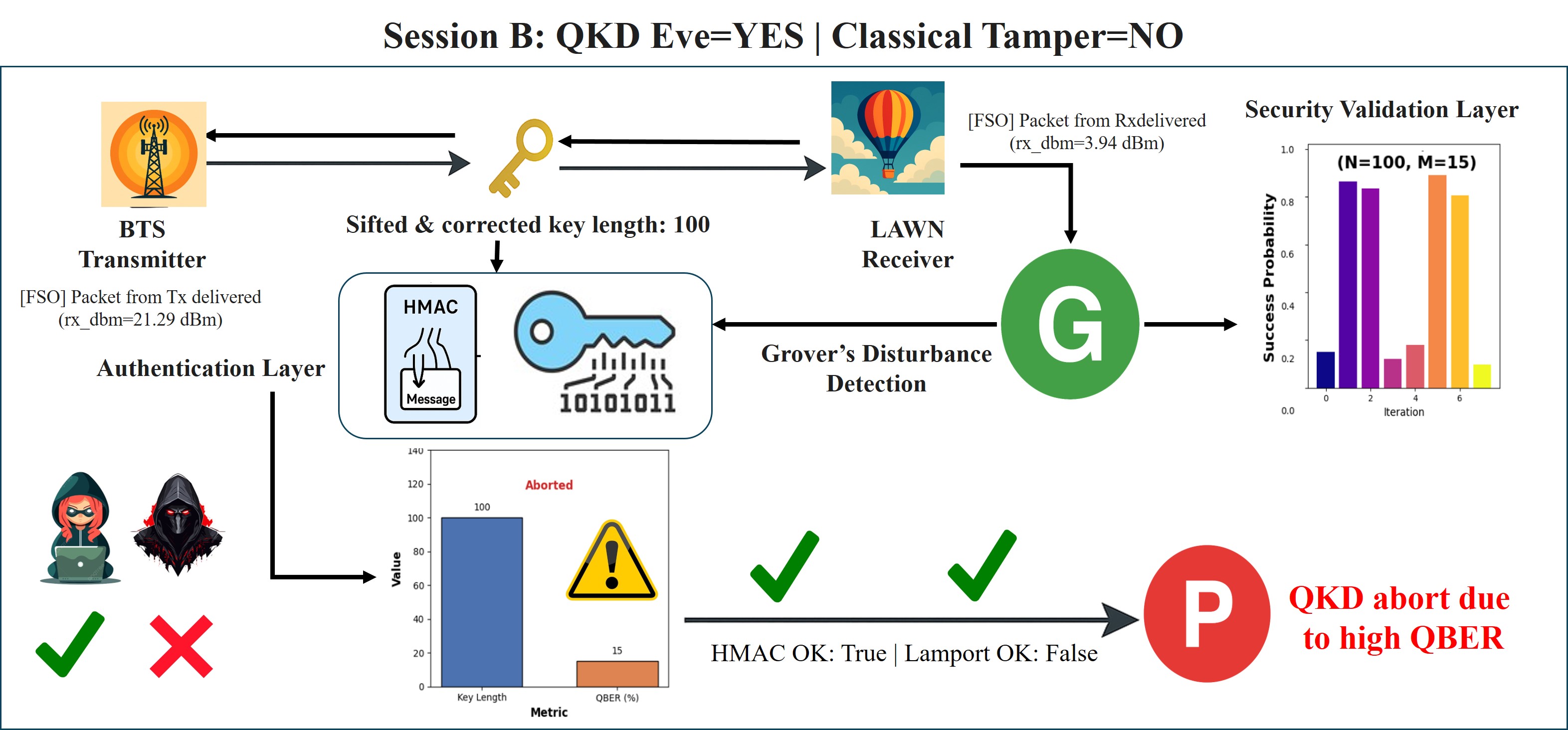}\\
    (c)\includegraphics[width=.46\textwidth]{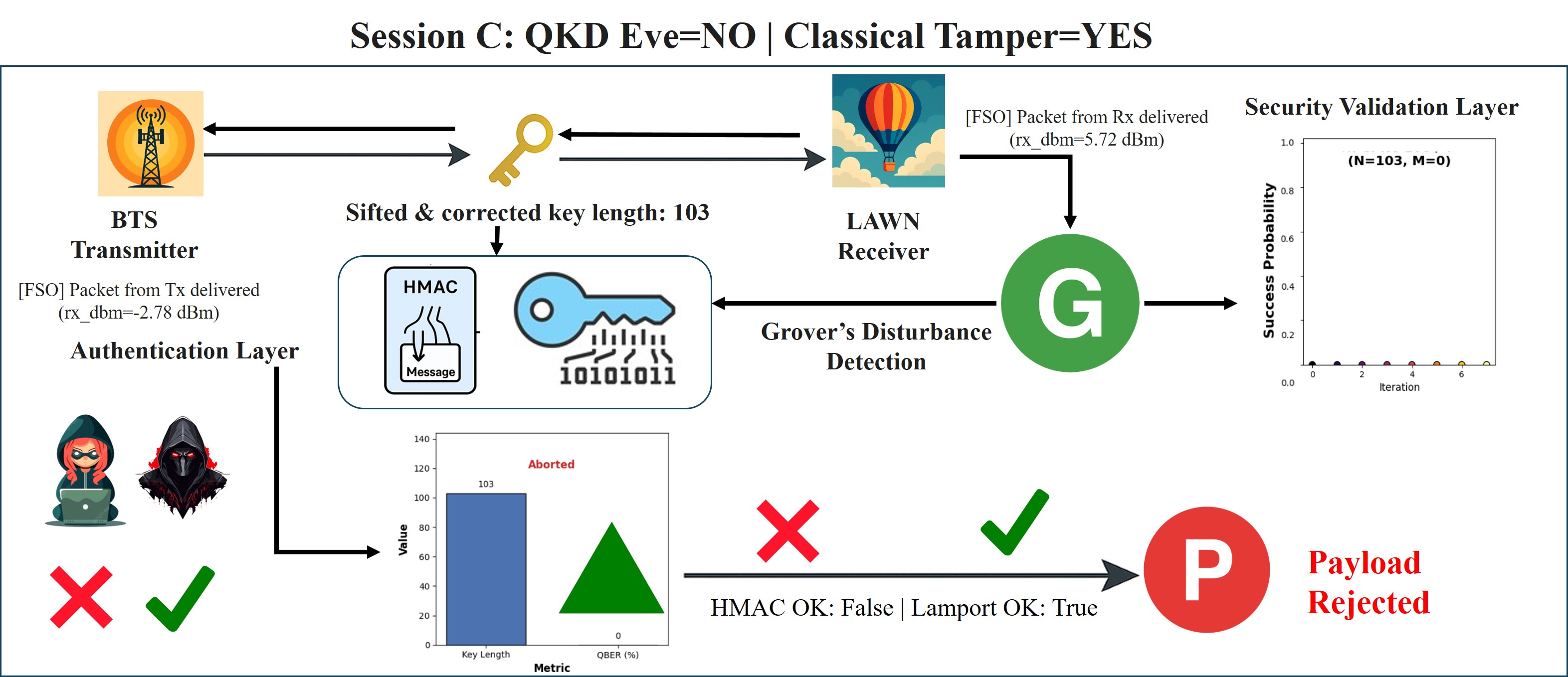}
    (d)\includegraphics[width=.46\textwidth]{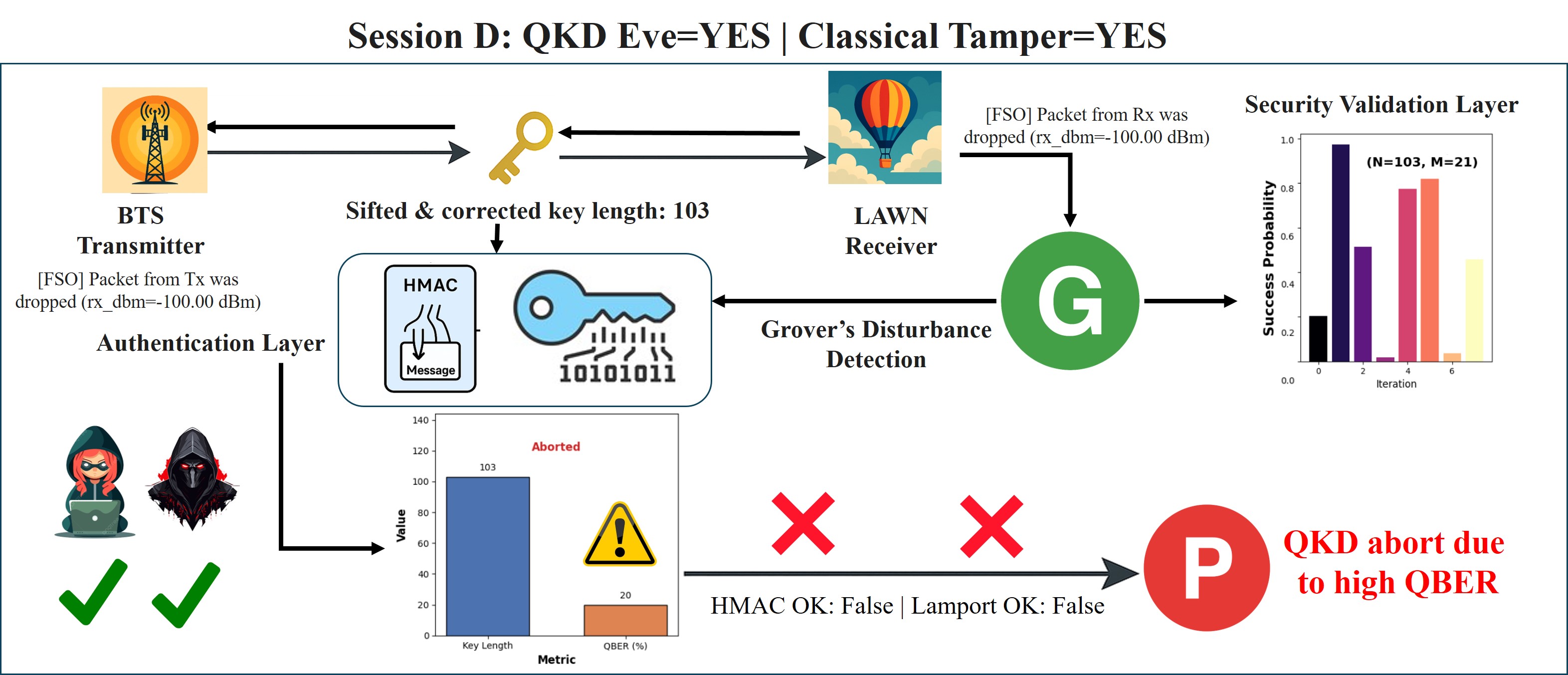} 
    \caption{Session Scenarios (a) A: QKD Eve=NO $|$ Classical Tamper=NO (b) B:QKD Eve=YES $|$ Classical Tamper=NO (c) C: QKD Eve=NO $|$ Classical Tamper=YES (d). D: Session: QKD Eve=YES $|$ Classical Tamper=YES }
    \label{fig:s}
\vspace{-0.5cm}
\end{figure*} 

\section{Session Scenarios: Security and Trust Implications}
To validate the effectiveness of the proposed architecture under operationally realistic conditions, we examine four representative communication sessions. Each scenario highlights distinct challenges encountered in BTS-LAWN communication, including quantum-layer errors, classical tampering, or both. This evaluation framework is tailored to airborne platforms, where environmental volatility and adversarial threats can compromise confidentiality, authenticity, and reliability.

\subsubsection{Session A: Nominal Operation with Valid Key and Authentication}
Fig.~\ref{fig:s}(a) and Fig.~\ref{fig:4}(a-i) illustrate a secure session under ideal channel conditions. The FSO link remains stable, with negligible turbulence or pointing error. QKD is completed successfully, yielding a sifted key with a zero QBER. The 98-bit session key is authenticated using Lamport OTS and validated through HMAC. Both checks succeed, confirming message integrity and sender authenticity. This scenario establishes a baseline for the system’s optimal behavior in low-disturbance environments.

\subsubsection{Session B: Quantum-Layer Interference Causing Session Abort}
In this case, as shown in Fig.~\ref{fig:s}(b) and Fig.~\ref{fig:4}(a-ii), the FSO channel is degraded due to turbulence or a potential quantum attack, such as an intercept-resend attack. The resulting QBER rises to 15\%, exceeding the secure threshold of 11\%. Consequently, the QKD session is aborted, and no session key is extracted. This response demonstrates the system's capacity to autonomously identify quantum-layer compromise and prevent downstream authentication, a critical safeguard for LAWNs exposed to rapidly changing atmospheric conditions or optical jamming.

\subsubsection{Session C: Classical Tampering Despite Secure Quantum Exchange}
Fig.~\ref{fig:s}(c) and Fig.~\ref{fig:4}(a-iii) present a session in which the quantum channel remains uncompromised, producing a 103-bit session key with zero QBER. However, the classical authentication layer is targeted at payload tampering. Verification fails at the LAWN, either due to an invalid Lamport signature or an incorrect HMAC, resulting in the rejection of the message. This highlights that even with secure key exchange, classical-layer integrity must be independently enforced, especially in LAWN operations where telemetry may be intercepted or spoofed.

\subsubsection{Session D: Compound Quantum and Classical Interference}
\begin{figure*}[h]
    \centering
    (a)\includegraphics[width=.90\textwidth]{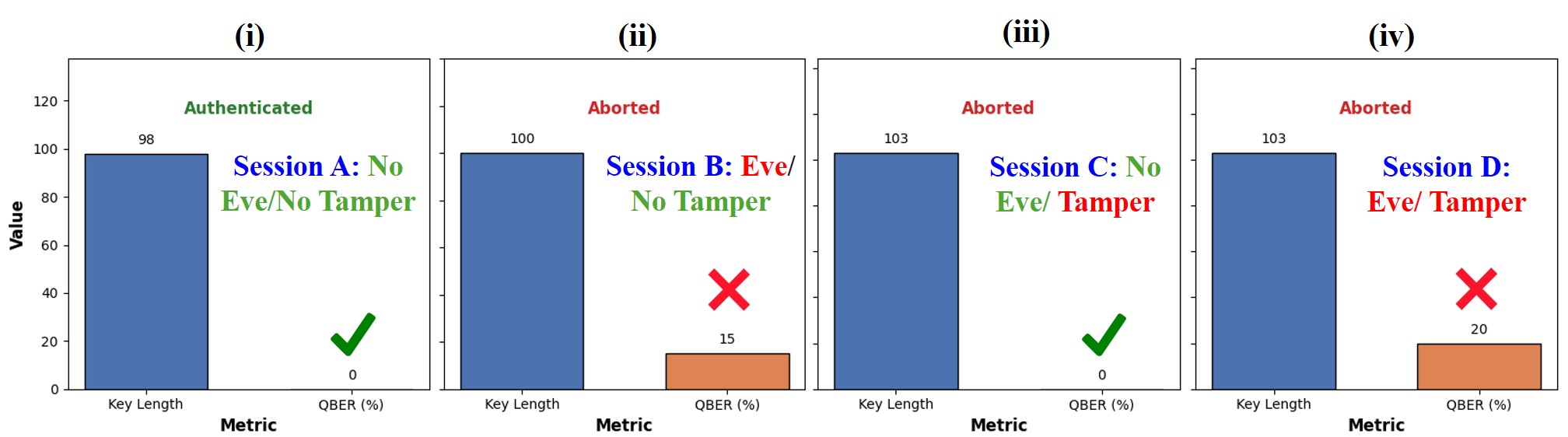}\\
    (b)\includegraphics[width=.80\textwidth]{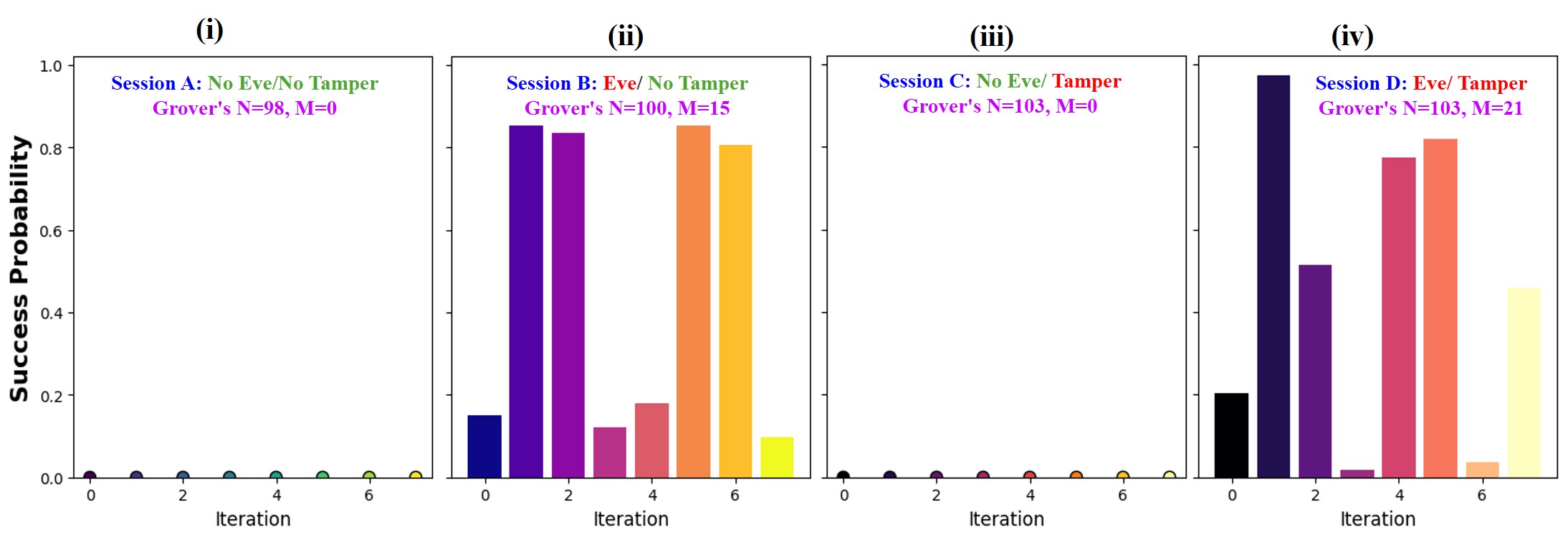}
    \caption{Session Scenarios (a) Security and trust implications (b) Grover’s Search Resilience Based on Session Outcomes.}
    \label{fig:4}
\vspace{-0.3cm}
\end{figure*}

As shown in Fig.~\ref{fig:s}(d) and Fig.~\ref{fig:4}(a-iv), this scenario combines disturbances at both communication layers. The QBER reaches 20\%, and the classical payload is corrupted during transmission. Neither key generation nor authentication succeeds. The session is aborted without accepting any message or key. This comprehensive failure case illustrates the architecture's layered defense model, ensuring that cross-domain interference does not propagate undetected. Such robustness is essential for LAWN deployments in contested airspaces or electronic warfare environments. 

These scenarios collectively demonstrate that reliable communication is permitted only when both quantum confidentiality and classical message integrity are satisfied. If either layer is compromised, the system aborts securely. This behavior aligns with the stringent trust requirements of LAWN missions and supports the deployment of resilient post-quantum communication frameworks in adversarial or degraded environments.

\vspace{-0.3cm}

\begin{table*}[h]
\tiny
\setlength{\tabcolsep}{3pt}
\centering
\caption{Comparative Analysis of Proposed Work with Quantum-Secure Architectures.}
\begin{tabular}{@{}lcccccccc@{}}
\toprule
\textbf{Metric} & \textbf{Proposed Work (2025)} & \textbf{\makecell{~\cite{zhao2020} (2020)}} & \textbf{\makecell{~\cite{alshaer2021} (2021)}} & \textbf{\makecell{~\cite{alshaer2022} (2022)}} & \textbf{\makecell{~\cite{nguyen2023} (2023)}} & \textbf{\makecell{~\cite{chehimi2024} (2025)}} & \textbf{\makecell{~\cite{wu2024} (2024)}} & \textbf{\makecell{~\cite{nguyen2024} (2024)}} \\
\midrule
QKD Protocol            & BB84+Lamport OTS +HMAC+ Grover        & CV-QKD                  & Entanglement-based (Eb)      & CV-QKD           & CV-QKD                & Eb+ RIS & MDI-QKD (OAM) & BB84 (with LDPC) \\
Channel Model           & Gamma-Gamma + weather & Fading model           & Static FSO              & Atmospheric + pointing error & Atmospheric + CV noise & Atmospheric loss + RIS & Turbulence + SDD + jitter & FSO turbulence (blind) \\
QBER Handling           & Grover + Dynamic     & Passive threshold       & Passive threshold       & Passive with threshold & Passive threshold     & None & Atmospheric modeling & Blind reconciliation \\
Auth Scheme             & Lamport OTS + HMAC   & RSA                     & RSA/ECC                 & Not used                    & ECC                   & Not used & Not specified & Not specified \\
Session Key Size        & 128-bit              & N/A                      & $\sim$100-bit           & N/A                & N/A        & N/A & N/A & N/A \\
Adversary Detection     & Real-time (Grover)   & None                    & None                    & None                   & None                  & None & None & None \\
Weather-Aware Eval.     & All types included   & Partial                 & No                      & Yes                    & Yes                   & Yes & Yes & Yes \\
Abort Strategy          & Quantum + Classical  & No                      & Quantum only            & Quantum only           & Quantum only          & None & Based on SDD limit & LDPC error bound \\
Trust Scoring           & Grover-based         & None                    & None                    & None                   & None                  & None & None & None \\
\textbf{Result/Effectiveness} & \makecell{High: secure \\ real-time threat detect.} & \makecell{Low: abort-prone \\ no detection} & \makecell{Medium: basic trust \\ no real-time check} & \makecell{Medium: QBER-only \\ abort} & \makecell{Medium: quantum-only \\ check} & \makecell{Low: no layered \\ validation} & \makecell{Medium: accounts for \\ jitter + SDD limits} & \makecell{Medium: blind reconciliation \\ improves key integrity} \\
\bottomrule
\end{tabular}
\label{tab:comparison}
\vspace{-0.5cm}
\end{table*}
\section{Grover-Based Anomaly Detection: Evaluating System Trustworthiness Under Attack}
While QBER alone provides a passive indication of quantum-layer anomalies, we extend the analysis using Grover’s quantum search model to assess how efficiently an intelligent adversary or a security auditor might locate compromised bits within a sifted key. In this context, the sifted key length \(N\) represents the size of the quantum search space, and the number of erroneous bits introduced by eavesdropping is denoted \(M\), derived as \(M = \mathrm{round}(\text{QBER} \times N)\). Grover’s algorithm provides a probabilistic advantage in locating one of these \(M\) "marked" items in \(O(\sqrt{N/M})\) queries. Our simulation evaluates Grover’s detection performance across the four session scenarios introduced earlier. Fig.~\ref{fig:4}(b) summarizes the computed success probabilities $\sin^2\bigl((2k+1)\,\arcsin\sqrt{M/N}\bigr)$ for iterations \(k=0\)-\(7\), across all four session.

\subsubsection{Session A: Clean Channel With No Errors}
In sessions where both quantum and classical layers are uncompromised (e.g., no eavesdropping or tampering), the QBER is exactly 0\%, yielding \(M = 0\) for key lengths around \(N = 103\) as shown in Fig.~\ref{fig:4}(b-i). As Grover’s algorithm relies on the presence of at least one marked item, the success probability remains zero across all iterations. This confirms the system's integrity and indicates no spurious detection under ideal conditions, a necessary baseline for trust.

\subsubsection{Session  B: Quantum-Layer Disturbance Detected}
In the quantum attack scenario, QBER rises to 15\%, producing approximately \(M = 15\) erroneous bits in a key of length \(N = 100\) as shown in Fig.~\ref{fig:4}(b-ii). Grover’s detection probability reaches 0.8640 after just one iteration and peaks at 0.8902 by the fifth iteration. These results demonstrate that Grover’s model is highly effective at flagging potential quantum layer interference early. For LAWNs, this means that intrusion attempts, such as intercept-resend or photon-number splitting, can be detected rapidly within milliseconds before authentication steps proceed.

\subsubsection{Session  C: Classical-Layer Tampering Without Quantum Errors}
This scenario examines a session in which the quantum exchange completes cleanly, producing a key of length \(N = 103\) with QBER = 0\%, meaning \(M = 0\) as shown in Fig.~\ref{fig:4}(b-iii). Despite the absence of quantum-layer disturbance, a classical-layer adversary modifies the signed payload or alters the transmitted Lamport signature or HMAC. Since Grover’s algorithm targets quantum-layer anomalies, its success probability remains zero across all iterations. This confirms that Grover is insensitive to classical tampering and further underscores the need for complementary HMAC and OTS validation layers. In LAWN deployments, such attacks could manifest as injected control messages or spoofed telemetry. The architecture detects them using lightweight hash-based checks, while Grover continues to serve as a quantum-layer integrity monitor.

\subsubsection{Session  D: Compound Quantum and Classical Attacks}
The most adversarial session involves both eavesdropping and tampering at the classical layer. Here, the QBER reaches 20\%, yielding \(M = 21\) errors in a key of length \(N = 103\) as shown in Fig.~\ref{fig:4}(b-iv). Remarkably, Grover’s success probability jumps to 0.9729 after a single iteration, confirming that even modest QBER rates make attacks highly detectable. The LAWN’s security module can therefore terminate the session before compromised keys are used, reinforcing the system’s proactive defense capability in high-risk zones.

\subsubsection{Interpreting Grover Oscillations and Iteration Limits}
Grover’s success probabilities exhibit oscillatory patterns, a phenomenon expected to occur due to \textit{over-rotation}. For example, after peaking at iteration 5, the likelihood in Session B drops slightly to 0.8354 at iteration 6. This behavior suggests that practical deployments should cap Grover iterations to the first local maximum, maximizing anomaly detection efficiency without unnecessary overhead.

\subsubsection{Implications for Secure NTN-QKD Integration}
Grover-based analysis provides a dynamic method to assess quantum-layer errors in NTN QKD systems. By mapping QBER to actionable threat indicators (\(M\)) and evaluating detection probability over key lengths (\(N\)), it enables lightweight, quantum-native verification ideal for resource-limited NTN nodes requiring fast decisions. Embedding Grover-based checks allows adaptive session management, where QKD sessions are aborted or restarted based on threat confidence. For sessions with nonzero QBER, Grover’s algorithm detects erroneous bits with high probability in just \(1\)–\(3\) iterations. When \(M = 0\), the success probability remains negligible, ensuring no false alarms and confirming clean sessions. The optimal number of iterations depends on the ratio \(M/N\), allowing intelligent networks to tune Grover’s run length for fast, efficient detection. This complements standard QBER checks, offering rapid response to moderate disturbances while avoiding unnecessary overhead during severe errors. While HMAC and one-time signatures protect the classical layer, Grover’s detection adds a quantum-native validation path. In cases of suspected tampering, a short Grover run can locate bit-level corruption, aiding selective retransmission or extra error correction,  thus reinforcing classical integrity mechanisms.

 Table~\ref{tab:comparison} demonstrates the superiority of our proposed architecture over existing quantum-secure NTN communication frameworks. Unlike prior works, which rely on passive QBER thresholds and lack adversarial awareness, our design integrates Grover dynamic error evaluation and trust scoring to enable real-time detection of quantum layer threats. Recent studies highlight that LAWN jitter, turbulence, and diffraction effects severely degrade QKD reliability \cite{wu2024}, while blind reconciliation is critical for dynamic links lacking prior error estimation \cite{nguyen2024}. To address these, we replace computationally heavy RSA and ECC schemes with lightweight Lamport OTS and HMACs, making the protocol suitable for resource constrained LAWNs. The inclusion of comprehensive weather-aware modeling, hybrid abort strategies, and end-to-end authentication further establishes our framework as a scalable, quantum resilient solution tailored for dynamic airborne environments. Also, this Grover informed evaluation complements QBER monitoring and enhances situational awareness in NTN quantum communication. Unlike static abort thresholds, Grover analytics provide real-time, scalable insight into potential compromise, offering intelligent trust assessment at the quantum layer.
\begin{figure*}
    \includegraphics[width=1\textwidth]{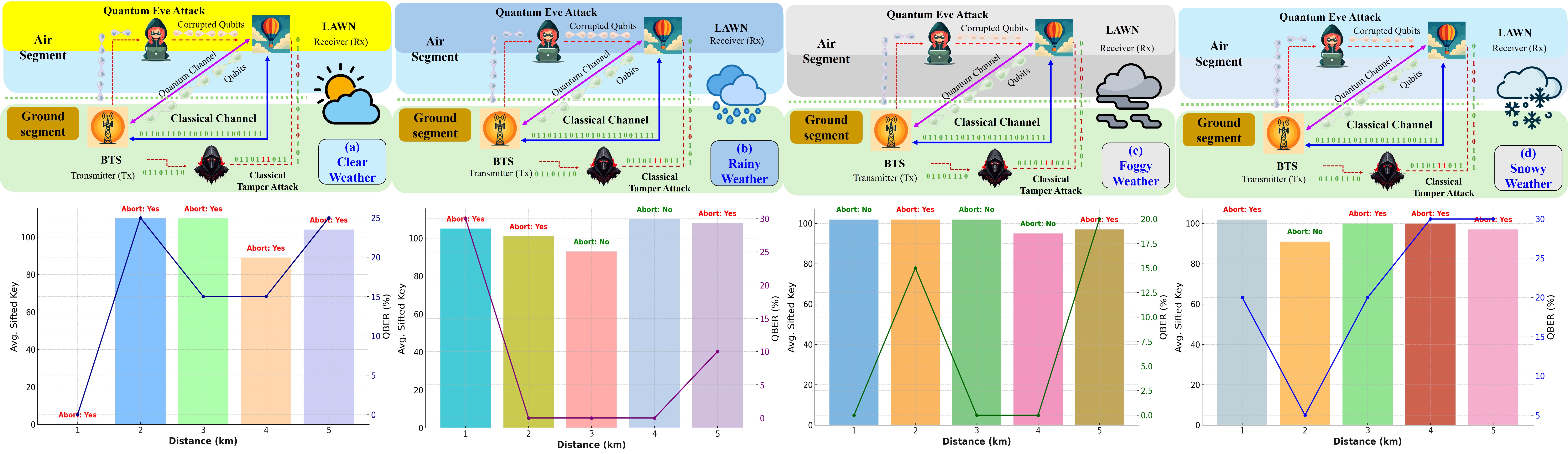} 
    \caption{Network Performance under Varying Distance and Weather Conditions.}
    \label{fig:5}
\end{figure*}  

The performance is further verified under different distances and weather conditions. For instance in Fig.~\ref{fig:5}, we present empirical results for the proposed network across varying link distances (1–5 km) under four distinct weather conditions: clear, fog, rain, and snow. Key metrics include average sifted key length, QBER, session abort frequency, and the success of payload authentication. These indicators offer insight into how environmental variability influences adaptive control and trust in intelligent quantum communication networks.

Despite expectations of optimal performance, clear-weather scenarios revealed inconsistencies. Notably, a complete lack of sifted key generation was observed at 1 km, paired with high abort rates across all distances. QBER values fluctuated between 0\% and 25\%, and payload authentication succeeded only at the shortest distance. These findings highlight the importance of real-time link diagnostics and the need to treat even “ideal” conditions as potentially unreliable for secure quantum operations. Fog-dominated environments demonstrated surprising robustness. Across all distances, average sifted key values remained steady (~95–102), and QBER was consistently low (0–20\%). Abort rates were relatively low, and authentication succeeded at a distance of 1 km. These conditions suggest fog does not significantly degrade quantum link fidelity at short to moderate ranges, supporting its inclusion in trusted operational windows for secure communication. Rain scenarios exhibited high variability: while QBER was acceptably low (0–10\%) at mid-range distances (2–4 km), frequent aborts and zero authentication successes indicate weaknesses at the classical verification layer. In snowy conditions, performance degraded more noticeably. QBER often exceeded 20\%, and abort rates were high across all distances. Although some sifted key exchange occurred, the persistent high QBER and authentication failures suggest that snow imposes substantial constraints on the reliability of quantum networks. Abort events occurred consistently whenever QBER surpassed tolerable thresholds. This strict binary behavior, while protective, limits the availability of links. Incorporating dynamic abort thresholds based on QBER trends, environmental context, and the application of advanced error correction could expand the trusted operational envelope. Such enhancements would enable intelligent quantum networks to better balance security guarantees with sustained availability under diverse environmental stressors.
\vspace{-0.3cm}
\section{Challenges, Bottlenecks, and Future Directions}
Despite advances in quantum-secure LAWN communication, key challenges remain for scalable, resilient, and real-time trust-aware systems in dynamic and adversarial airborne environments.

\subsubsection{Environmental and Physical Layer Limitations}
LAWN–BTS quantum links offer lightweight post-quantum communication but remain vulnerable to weather. FSO channels degrade under fog, rain, snow, and solar glare, raising QBER and causing photon loss. LAWN mobility introduces beam misalignment and jitter. While filtering helps, adaptive optics and accurate beam steering are still underdeveloped. Similar vulnerabilities exist for HAP and LEO platforms, which also suffer from orbital jitter and space weather effects.

\subsubsection{Quantum Storage and Synchronization Constraints}
Photon qubits decohere rapidly in turbulence. LAWNs lack quantum memory and must process qubits in sub-second windows. Synchronizing quantum and classical layers is difficult due to latency, motion, and clock drift. These issues are amplified in LEO/HAP due to longer link distances and Doppler effects, complicating timing and alignment.

\subsubsection{Post-Quantum Cryptographic and Computational Overheads}
Even lightweight PQC schemes like Lamport OTS impose energy and processing burdens. Each signature may require hundreds of SHA-256 hashes, taxing onboard resources. Lattice alternatives, though efficient, still demand careful trade-offs in power-constrained platforms like LAWNs or small LEO payloads.

\subsubsection{Cross-Layer and Network-Level Limitations}
Current routing lacks awareness of quantum-layer constraints such as entanglement fidelity or photon budget. Efficient routing for NTN must include quantum metrics, especially for dynamic topologies with multiple LEO/HAP relays. Lack of quantum-aware switching and buffering restricts scalability and load balancing.

\subsubsection{Security Under Partial Compromise and Adversarial Threats}
Hybrid systems, though secure in theory, remain exposed to side-channels, leakage, and denial-of-service attacks on classical telemetry. Even with QKD, tampering or spoofing can degrade trust. LEO-based relays are especially at risk due to remote access and a lack of local verification. Strong intrusion detection and fallback protocols are needed.

\subsubsection{Dynamic Adaptation and Operational Constraints}
QBER-triggered aborts help preserve security but can interrupt time-critical operations. Future protocols must adapt dynamically, balancing security with availability, especially during high-mobility or emergency missions. For LEO, delay-tolerant mechanisms are crucial due to intermittent ground contact.

\subsubsection{Machine Learning and Predictive Trust Scoring}
Compact machine learning models can predict QBER trends using weather, mobility, and link data. Such tools support preemptive adaptation and error mitigation. Combined quantum-classical trust scores will help nodes autonomously reroute or abort when compromise is likely.

\subsubsection{Toward Scalable Quantum Mesh Networks}
Transitioning from point-to-point to mesh topologies will improve reliability. Coordinated control among LAWNs, HAPs, and LEO satellites requires cross-layer routing and shared trust models. Testbeds are essential for validating Grover-based detection, adaptive optics, and dynamic key refresh. Bridging theoretical security and real-time operation is key to deployment at scale.

\vspace{-0.3cm}
\section{Conclusion}
The increasing use of low-altitude wireless networks (LAWN) in both civilian and defense applications demands secure communication systems that are resilient to both classical and quantum threats. This work presented a hybrid architecture combining BB84-based quantum-key distribution over free space optics links with lightweight post-quantum authentication using Lamport one-time signatures. The proposed system enables real-time detection of eavesdropping, aborts compromised sessions, and secures classical messages against tampering, even under adverse weather conditions. Simulations confirm robust key generation under moderate conditions, with Grover-based anomaly detection adding a quantum-native layer for rapid threat identification. To further enhance resilience, future research should explore lightweight machine learning for adaptive quantum bit error rate management and consider integrating lattice-based signatures for stronger classical-layer security. This layered framework offers a practical and scalable solution for secure LAWN communication in the quantum era.
\vspace{-0.3cm}

\bibliographystyle{IEEEtran}
\bibliography{cas}

\begin{thebibliography}{10}
\providecommand{\url}[1]{#1}
\csname url@samestyle\endcsname
\providecommand{\newblock}{\relax}
\providecommand{\bibinfo}[2]{#2}
\providecommand{\BIBentrySTDinterwordspacing}{\spaceskip=0pt\relax}
\providecommand{\BIBentryALTinterwordstretchfactor}{4}
\providecommand{\BIBentryALTinterwordspacing}{\spaceskip=\fontdimen2\font plus
\BIBentryALTinterwordstretchfactor\fontdimen3\font minus \fontdimen4\font\relax}
\providecommand{\BIBforeignlanguage}[2]{{%
\expandafter\ifx\csname l@#1\endcsname\relax
\typeout{** WARNING: IEEEtran.bst: No hyphenation pattern has been}%
\typeout{** loaded for the language `#1'. Using the pattern for}%
\typeout{** the default language instead.}%
\else
\language=\csname l@#1\endcsname
\fi
#2}}
\providecommand{\BIBdecl}{\relax}
\BIBdecl

\bibitem{sharma2021}
P.~Sharma, A.~Agrawal, V.~Bhatia, S.~Prakash, and A.~K. Mishra, ``Quantum key distribution secured optical networks: A survey,'' \emph{IEEE Open Journal of the Communications Society}, vol.~2, pp. 2049--2083, 2021.

\bibitem{alshaer2021}
N.~Alshaer, A.~Moawad, and T.~Ismail, ``{Reliability and Security Analysis of An Entanglement Based QKD Protocol in A Dynamic Ground-to-UAV FSO Communications System},'' \emph{IEEE Access}, vol.~9, pp. 168\,052--168\,067, 2021.

\bibitem{sinha2023}
S.~Sinha and C.~Kumar, ``Performance evaluation of uav-assisted fso link in generalized malaga distributed atmospheric turbulence conditions,'' \emph{Optical and Quantum Electronics}, vol.~55, no.~13, p. 1161, 2023.

\bibitem{guo2022}
W.~Guo, Y.~Zhan, T.~A. Tsiftsis, and L.~Yang, ``{Performance and channel modeling optimization for hovering UAV-assisted FSO links},'' \emph{Journal of Lightwave Technology}, vol.~40, no.~15, pp. 4999--5012, 2022.

\bibitem{nguyen2023}
T.~V. Nguyen, H.~T. Le, H.~Pham, V.~Mai, and N.~Dang, ``{Enhancing Design and Performance Analysis of Satellite Entanglement-Based CV-QKD/FSO Systems},'' \emph{IEEE Access}, vol.~11, pp. 112\,097--112\,107, 2023.

\bibitem{sharma2024}
P.~Sharma, D.~Singh, and S.~Ramabadran, ``{Performance Analysis of UAV-Based FSO Communication over Doubly Inverted Gamma-Gamma Turbulence Channel},'' in \emph{2024 National Conference on Communications (NCC)}, 2024, pp. 1--6.

\bibitem{lu2022}
K.-A. Shim, ``A survey on post-quantum public-key signature schemes for secure vehicular communications,'' \emph{IEEE Transactions on Intelligent Transportation Systems}, vol.~23, no.~9, pp. 14\,025--14\,042, 2021.

\bibitem{ramabadran2023}
A.~Kumar, S.~Bhatia, K.~Kaushik, S.~M. Gandhi, S.~G. Devi, D.~A. D.~J. Pacheco, and A.~Mashat, ``Survey of promising technologies for quantum drones and networks,'' \emph{IEEE Access}, vol.~9, pp. 125\,868--125\,911, 2021.

\bibitem{alallaq2024}
Z.~J. Al-Allaq and W.~M.~R. Shakir, ``{A Comprehensive Analysis of FSO Communications with UAV Technologies},'' in \emph{2024 International Symposium on Networks, Computers and Communications (ISNCC)}, 2024, pp. 1--7.

\bibitem{gupta2023}
A.~Gupta, D.~Dhawan, and N.~Gupta, ``{Review on UAV-based FSO links: recent advances, challenges, and performance metrics},'' \emph{Optical Engineering}, vol.~63, no.~4, p. 041204, 2024.

\bibitem{chehimi2024}
M.~Chehimi, M.~Elhattab, W.~Saad, G.~Vardoyan, N.~K. Panigrahy, C.~Assi, and D.~Towsley, ``Reconfigurable intelligent surface (ris)-assisted entanglement distribution in fso quantum networks,'' \emph{IEEE Transactions on Wireless Communications}, 2025.

\bibitem{alshaer2022}
N.~Alshaer and T.~Ismail, ``{Performance Evaluation and Security Analysis of UAV-Based FSO/CV-QKD System Employing DP-QPSK/CD},'' \emph{IEEE Photonics Journal}, vol.~14, no.~3, pp. 1--11, 2022.

\bibitem{zhao2020}
H.~Zhao and M.-S. Alouini, ``On the transmission probabilities in quantum key distribution systems over fso links,'' \emph{IEEE Transactions on Communications}, vol.~69, no.~1, pp. 429--442, 2020.

\bibitem{wu2024}
D.~Wu, J.~Li, L.~Yang, Z.~Deng, J.~Tang, Y.~Cao, Y.~Liu, H.~Hu, Y.~Wang, H.~Yu \emph{et~al.}, ``Practical performance analysis of mdi-qkd with orbital angular momentum on uav relay platform,'' \emph{Entropy}, vol.~26, no.~8, p. 635, 2024.

\bibitem{nguyen2024}
C.~T. Nguyen, H.~D. Le, V.~V. Mai, P.~V. Trinh, and A.~T. Pham, ``Blind reconciliation with protograph ldpc code extension for fso-based satellite qkd systems,'' in \emph{2024 14th International Symposium on Communication Systems, Networks and Digital Signal Processing (CSNDSP)}.\hskip 1em plus 0.5em minus 0.4em\relax IEEE, 2024, pp. 17--22.

\end{thebibliography}
\begin{IEEEbiographynophoto}
{Zeeshan Kaleem [Senior Member IEEE]} is serving as an Assistant Professor in King Fahd University of Petroleum and Minerals. 
\end{IEEEbiographynophoto} \vspace{-1.5cm}
\begin{IEEEbiographynophoto}{Misha Urooj Khan} is with European Organization for Nuclear Research, CERN, Switzerland.
\end{IEEEbiographynophoto} \vspace{-1.5cm}
\begin{IEEEbiographynophoto}{Ahmad Suleman} is with AITeC, National Center for Physics (NCP), Pakistan.
\end{IEEEbiographynophoto}\vspace{-1.5cm}
\begin{IEEEbiographynophoto}{Waqas Khalid} is with the Institute of Industrial Technology, Korea University.
\end{IEEEbiographynophoto}\vspace{-1.5cm}
\begin{IEEEbiographynophoto}{Kai-Kit Wong [IEEE Fellow]} is with the Department of Electronic and Electrical Engineering, University College London, United Kingdom, and also with Yonsei Frontier Lab, Yonsei University, Seoul, Korea.
\end{IEEEbiographynophoto}\vspace{-1.5cm}
\begin{IEEEbiographynophoto}{Chau Yuen [IEEE Fellow]} is with the School of Electrical and Electronics Engineering,
Nanyang Technological University, Singapore.
\end{IEEEbiographynophoto}\vspace{-1.5cm}

\end{document}